\documentclass{vgtc}                          

\graphicspath{{figures/}{pictures/}{images/}{./}} 
\usepackage{amsmath}
\usepackage{algpseudocode}
\usepackage{algorithm}
\usepackage{caption}
\usepackage{times}                     

\usepackage{tabu}                      
\usepackage{booktabs}                  
\usepackage{lipsum}                    
\usepackage{mwe}                       

\usepackage{mathptmx}                  
\newcommand\chang[1]{{\color{black}#1}}
\newcommand\kate[1]{{\color{black}#1}}

\onlineid{1156}

\vgtccategory{Research}

\vgtcinsertpkg

\title{An OverviewDetail Layout for Visualizing Compound Graphs}

\author{Chang Han\thanks{e-mail: changhan@sci.utah.edu}\\ %
        \scriptsize The University of Utah %
\and Justin Lieffers
\\%
     \scriptsize The University of Arizona %
\and Clayton Morrison\thanks{e-mail: clayton@lum.ai}\\%
\scriptsize Lum AI
\and Katherine E. Isaacs\thanks{e-mail: kisaacs@sci.utah.edu}\\ %
     \scriptsize The University of Utah 
}

\abstract{
Compound graphs are networks in which vertices can be grouped into larger subsets, with these subsets capable of further grouping, resulting in a nesting that can be many levels deep. \kate{In} several applications, including biological workflows, chemical equations, and computational data flow analysis, \kate{these graphs often exhibit a tree-like nesting structure, where sibling clusters are disjoint}. Common \kate{compound graph} layouts prioritize the lowest level of the grouping, down to the individual ungrouped vertices, which can make the higher level grouped structures more difficult to discern, especially in deeply nested networks. \kate{Leveraging the additional structure of the tree-like nesting, we} contribute an overview+detail layout \kate{for this class of compound graphs} that preserves the saliency of the higher level network structure when groups are expanded to show internal nested structure. Our layout draws inner structures adjacent to their parents, using a modified tree layout to place substructures. We describe our algorithm and then present case studies demonstrating the layout's utility to a domain expert working on data flow analysis. Finally, we discuss network parameters and analysis situations in which our layout is well suited.
} 


\begin{document}


\firstsection{Introduction}

\maketitle

Compound graphs are networks in which additional grouping information is available regarding the vertices. A group may contain other groups of vertices, leading to a hierarchical nesting of network contents~\cite{vehlow2017visualizing}. These networks with grouping structures arise frequently in multi-step processes. For example, vertices relating to a biological process may be grouped into a compound node representing that process~\cite{wu2021graph, gosak2018network}. Larger, more complicated processes may be composed of these compound nodes. Thus, the compound graph may have many levels of grouping to navigate. 

\begin{figure}[H]
    \centering
    \includegraphics[width=1\linewidth]{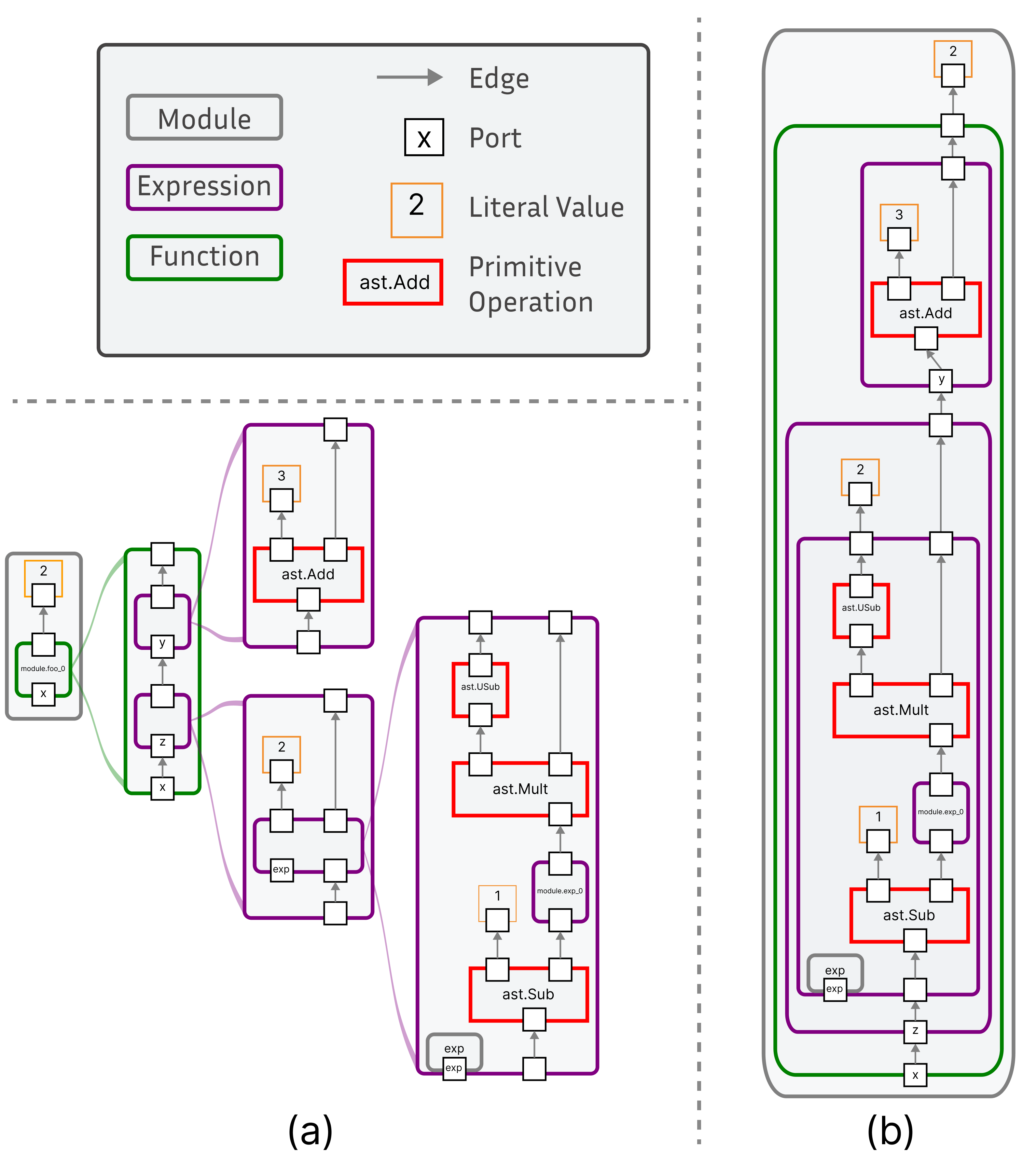}
    \caption{An overview+detail layout (a) vs. focus+context layout (b) of the same compound graph. The graph depicts the data flow of a small Python program. In our case study, a domain expert looks at compound networks of larger, more complicated programs.}
    \label{fig:teaser}
\end{figure}

A common way to visualize graphs is with a node-link diagram, which promotes tracing paths through the network. With compound graphs, the initial depiction of the graph may show only the higher level groupings, reducing clutter by allowing the user to interactively expand the groups to see their members~\cite{wongsuphasawat2017visualizing}. Visually, the groups are often expanded in place, resulting in a layout that focuses on lower levels but can obsucre higher level structure, as shown in \autoref{fig:teaser}(b).

\kate{For compound graphs where sibling clusters are disjoint, e.g., those with tree-like hierarchies like the multi-step process graphs discussed above, we can balance the display of both high level and low level structure through an overview+detail approach.}
We propose a layout (\autoref{fig:teaser}(a)) that perserves the higher level structure while still supporting examination of lower level structure, placing the two near each other so users can simultaneously explore both the internal details of a group and how that group interacts with its peers at the higher level. We use two strategies to achieve this layout. First, we explicitly route edges that traverse group membership through ``ports'' that isolate the inner group layout while preserving path tracing. Second, we adapt the Flexible RT algorithm~\cite{van2014drawing} to draw expanded groups near their collapsed representations within a group.

We present our contributed layout in \autoref{sec:technique}. We then demonstrate the utility of the layout through a collection of case studies in analyzing data flow graphs. We conclude with a discussion of situations in which our layout is appropriate. Specifically, we consider our layout most appropriate for directed compound networks that have multiple levels of nesting, especially when analysis tasks benefit from a strong understanding of higher level contexts along with expanded detail.

\section{Related Work}
\label{sec:relatedwork}
A common way to visualize compound graphs is with Sugiyama-style algorithms~\cite{eades1997straight, forster2002applying, sugiyama1991visualization, sander1996layout}.
Sugiyama and Misue~\cite{sugiyama1991visualization} initially proposed an algorithm to create a compound layout for directed graphs. Their method generates a layered layout within each cluster. Extending this approach, Sander~\cite{sander1996layout} introduced global partitioning and then drew borders around each cluster. Despite employing various techniques to reduce edge crossings, the layout becomes cluttered as the size of the graph increases. To mitigate this issue, Wongsuphasawat et al.\cite{wongsuphasawat2017visualizing} applied edge bundling to compound graphs, routing edges between clusters so that they connect only nodes that are siblings within the hierarchy. This strategy also aids in maintaining layout stability during interaction, such as expanding groups~\cite{misue1995layout}. However, edge bundling complicates the tracing of edges and the graph's appearance still changes significantly if an expanded group is large. Yoghourdjian et al.~\cite{yoghourdjian2015high} proposed a grid layout method, designed for small  compound graphs, that produces compact layouts with edges routed between grid cells. 

Deriving compound graphs, such as through clustering, is a common approach for visualizing overviews of large and complex undirected graphs. Balzer and Deussen~\cite{balzer2007level} proposed a level-of-detail technique for visualizing clustered graphs in both 2D and 3D. They employ implicit surfaces and edge bundling to simplify the visual representation of these graphs. Additionally, Ham and van Wijk~\cite{van2004interactive} proposed a method to interactively inspect local structures of small-world graphs, while maintaining a global overview of the entire structure. \chang{Abello et al.~\cite{Abello2006} proposed an interactive system for large scale compound graphs that allows users to navigate through the hierarchies by expanding clusters.} 
Archambault et al.~\cite{archambault2008grouseflocks} combined a tree view and a graph view to aid exploration of compound graphs. The tree view offers a clear representation of the graph hierarchy, while the graph view displays the current cut of the graph hierarchy, selected via the tree view. \chang{The same combined views~\cite{Archambault2010} is also used to modify graph hierarchy.} Our layout combines tree and graph notions, thereby providing both a clear hierarchy and the detailed graph structure within the same view. 

\section{Proposed Compound Graph Layout}
\label{sec:technique}

 The primary goal of the proposed layout is to preserve the original graph representations at each level of the hierarchy, thereby enabling viewers to understand the network at multiple levels simultaneously. Thus, we aim to have little-to-no distortion at each level, even when expanding down the hierarchy. Before introducing the algorithm, we first introduce the concept of \textit{ports} that we have employed when visualizing compound graphs.

\subsection{Ports and Inner Layout for Expanded Groups}
Often in compound graphs, edges crossing into a compound node either are routed directly to their neighbor within the compound node or are represented as terminating at the compound node boundary, with the explicit connections elided, often to remove clutter. For example, TensorFlow graphs~\cite{wongsuphasawat2017visualizing} bundle edges entering a compound node.

\begin{figure}
    \centering
    \includegraphics[width=1\linewidth]{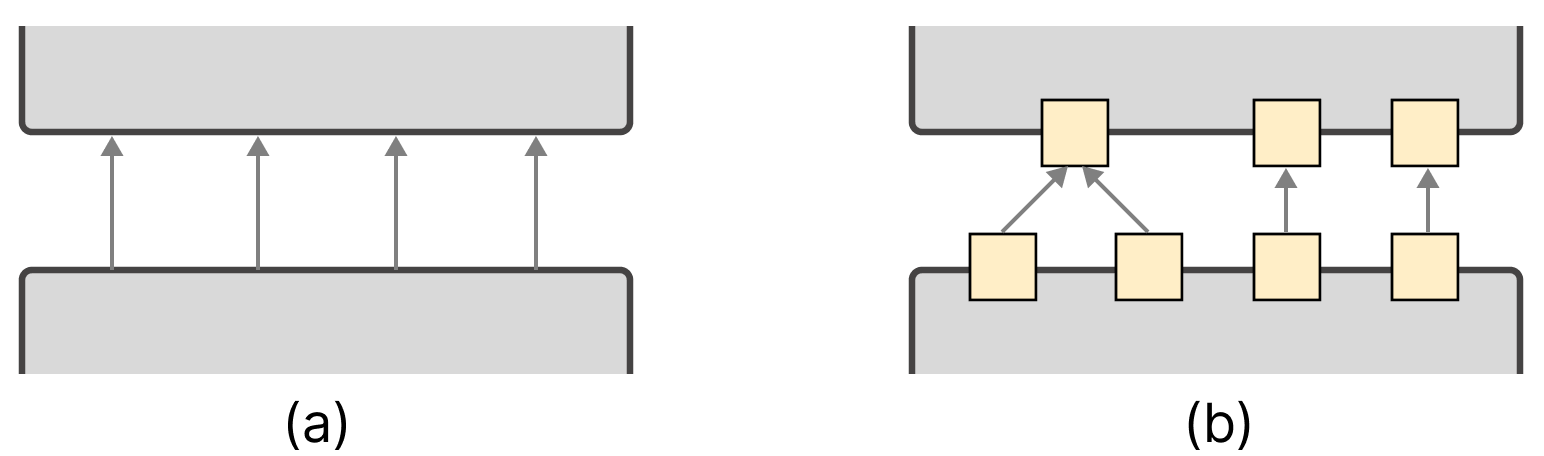}
    \captionsetup{belowskip=-10pt}
    \caption{\chang{Compared to (a), }we explicitly incorporate ports into the design in (b) to increase the clarity of links through compound graphs and isolate the the layout within compound nodes.}
    \label{fig:ports}
\end{figure}

Several layouts have used a concept of `ports' to guide the routing of edges. For example, in the {\em dot} algorithm implementation in GraphViz~\cite{ellson2002graphviz}, ports specify which side of a node the edge should be routed to.
Our layout draws explicit ports, shown in our figures as small squares (\autoref{fig:ports}). Edges crossing the border of a compound node are routed through ports for clarity, with one direction (``top'') indicating `in' ports and othe opposed direction (``bottom'') `out' ports. Our layout allows for any numer of ports per compound node. This choice allows us to layout the members of the compound node independently of its outer context, except for the direction in which the outer edges come (i.e., the direction of the ports).

Explicit boundary crossing ports has relevance in several domains. In computational graphs, operations (nodes) can receive data variables as inputs and produce new variables as outputs~\cite{wongsuphasawat2017visualizing}. Similarly, chemical equations may have similar input and output semantics~\cite{atkins2023atkins}. Note that our ports are agnostic to layout methods, meaning that the proposed technique is applicable to graphs both with and without an explicit presentation of ports.

We employ Dagre, a JavaScript library, to compute a Sugiyama-style layout for each expanded group~\cite{sugiyama1981methods}. We treat the 
ports as child nodes within the variable boxes and add dummy nodes between the input and output ports to separate them.

\subsection{Overview+Detail Metaphor for Compound Graphs}
The overview+detail design in visualization context refers to \textit{``the simultaneous display of both an overview and detailed view of an information space, each in a distinct presentation space''}~\cite{cockburn2009review}. Existing compound graph visualizations focus on ``focus+context''~\cite{schaffer1996navigating, gansner2005topological} and ``semantic zooming''~\cite{wongsuphasawat2017visualizing}, both which distort the higher levels around the focus level. The zooming techniques also cause the structure of the pre-zoomed graph (the overview) to be less salient, making it difficult to maintain a sense of higher level connectivity and structure when viewing the internals of compound nodes.

\begin{figure}
    \centering
    \includegraphics[width=1\linewidth]{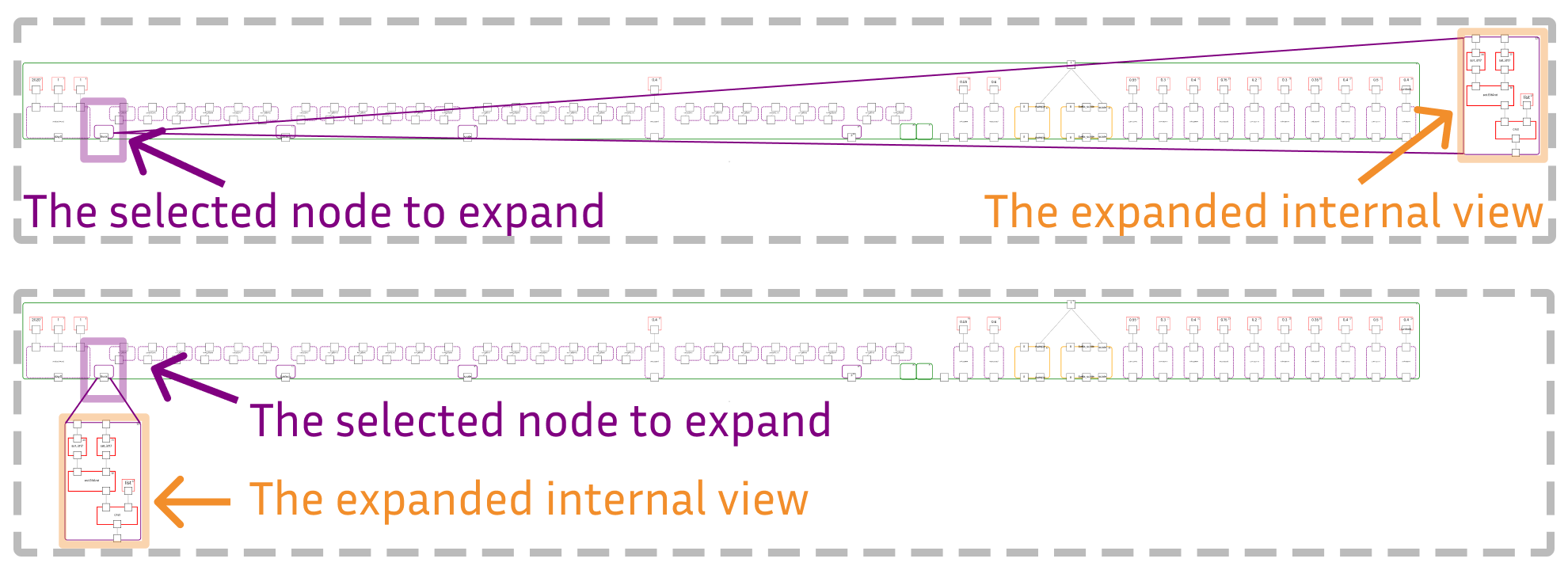}
    \caption{(Top) The Reingold-Tilford algorithm can produce poor results when the compound node is oblong in the direction of tree growth. Even growing the layout vertically, the internal view would be placed in the middle of the oblong compound node rather than near its counterpart. (Bottom) The ideal layout would place expanded nodes near their collapsed counterparts. }
    \label{fig:RTfailedcase}
\end{figure}

To address this issue, we propose a new layout for compound graphs based on the overview+detail metaphor in interface design. This layout creates a distinct view of the subgraph at each cluster expansion, instead of expanding selected node and re-laying out the graph containing it. The nesting relationship between the original graph and the subgraphs enables us to employ tree layout methods to arrange these distinct views. However, traditional tree layout algorithms like Reingold-Tilford (RT) algorithm~\cite{reingold1981tidier} cannot be directly applied for two main reasons: i) the boxes of different view frames have siginificant size and aspect ratio differences, and ii) placing child nodes at the center of one compound (parent) node direction, as is the RT default, may result in layouts where the collapsed and expanded representations are far from each other. Below we discuss our customized variant of the RT algorithm to address these two issues.

\subsection{Orthogonal Non-layered Reingold-Tilford Algorithm}

As shown in~\autoref{fig:RTfailedcase} (Top), simply applying RT algorithm can yield poor results. Placing the expanded node far from its collapsed counterpart creates visual edge clutter and presents difficulties in simultaneously comprehending the node in both its collapsed and expanded states. \chang{Although changing the tree layout direction in RT} can alleviate the issue in this case, it is not a viable solution because (1), even if centered vertically, the expanded node would still be some distance from its collapsed representation, and (2) the graph may contain elements that are optimized for left-to-right and top-to-bottom layouts seperately. Below we discuss our customized variant of RT, which mitigates these issues. \chang{An implementation of this algorithm can be found on Github\footnote{\url{https://github.com/ml4ai/moviz-client}}.}

The primary design goal of our method is to \textbf{minimize the distance between the selected node and its expanded view}. We base our algorithm on the ``Flexible Reingold-Tilford (Flexible RT)''algorithm~\cite{van2014drawing}, a non-layered variant of the traditional RT algorithm. Compared to the traditional RT, the Flexible RT achieves a higher space utilization rate and helps reduce edge length. We employ the non-layered RT because layers do not encode critical information in our use cases. For applications requiring a clear layer structure, traditional RT can be re-employed, albeit at the expense of space utilization and edge length.

\begin{figure}
    \centering
    \includegraphics[width=1\linewidth]{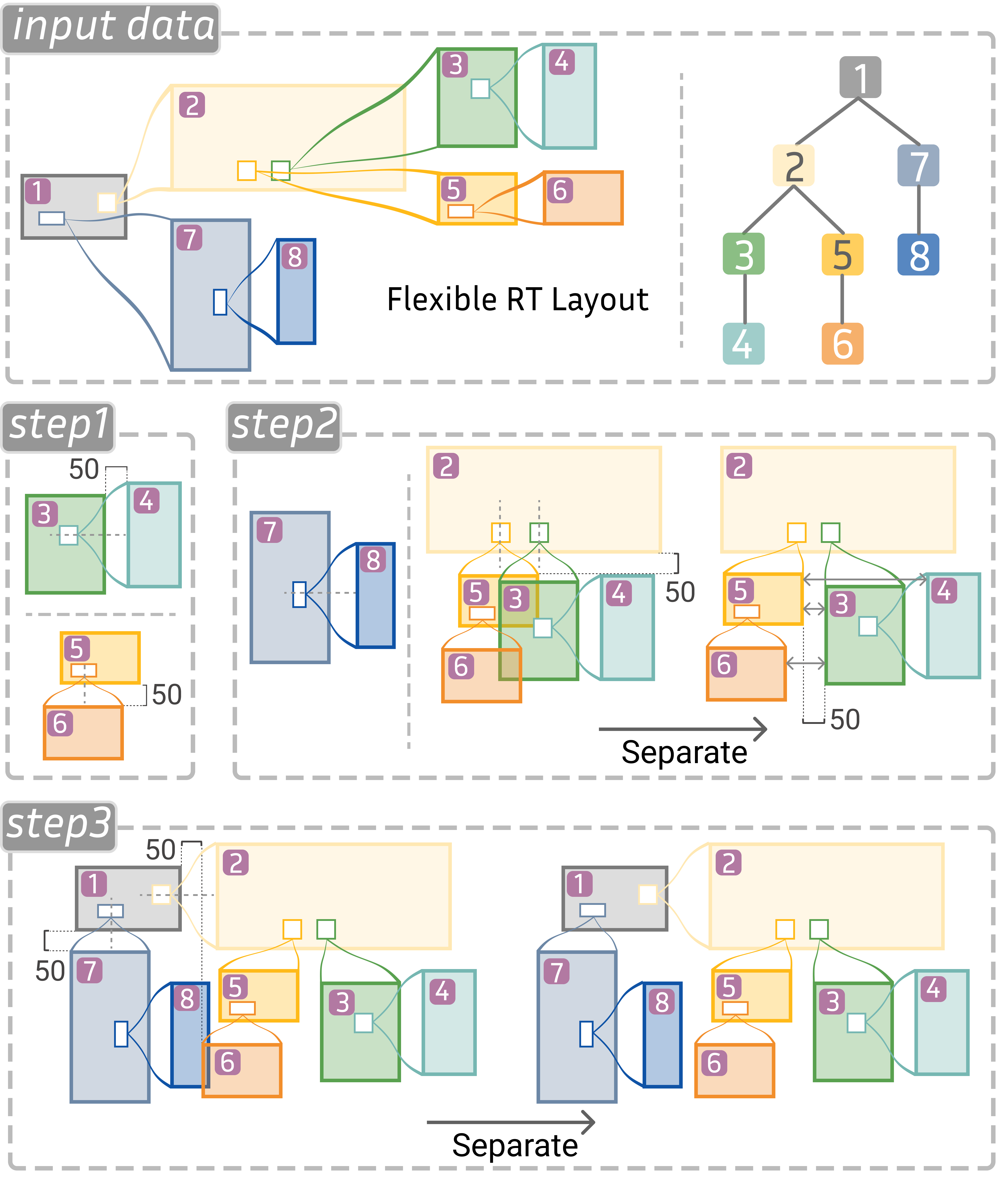}
    \caption{An illustration of our proposed variant of \chang{RT} algorithm. The input data is shown in both our layout and a simplified tree view. \chang{Then in each step, }we follow the RT bottom-up placement. \chang{We frist place group parents with respect to expanded children based on the position of their corresponding internal nodes, and then make separation passes in both directions of tree expansion.}}
    \label{fig:modifiedRT}
\end{figure}

Our proposed method adopts a bottom-up approach similar to both RT algorithms, as shown in Figure~\ref{fig:modifiedRT}. The primary distinction of our algorithm is that it lays out the tree in two directions. As illustrated in Figure~\ref{fig:modifiedRT} {\em step1}, the children (nodes 4 and 6) are positioned to the right and bottom, respectively, based on their proximity to the corresponding boundaries of the parent boxes (nodes 3 and 5). This strategy effectively reduces the distance between the node and its expanded view.

Another notable difference from previous methods is our initialization of each child node's position. At each step, we initialize the position of the child according to its location within the parent box. For instance, in {\em step2}, before separation, the x-coordinates of the two subtrees (3-4 and 5-6) are set to align with their location in the parent box. Following this initialization, the algorithm checks for overlaps and applies a separation operation 
as necessary, utilizing the same separation technique as described in the Flexible RT algorithm~\cite{van2014drawing}. Subsequently, the parent node adjusts to the average displacement of all child nodes, minimizing the total displacement of child nodes. 

Note that this separation is performed for child subtrees that are oriented in the same direction (bottom or right). In some scenarios, as shown in Figure~\ref{fig:modifiedRT} {\em step3} (left), the children oriented in the left-right direction may overlap with those in the top-down direction. Because the contours of these groups can form non-convex polygons, resolving their overlap becomes a problem of finding the Minimum Translation Vector (MTV) to separate overlapping non-convex geometric shapes, a challenge commonly addressed in collision detection within computer graphics~\cite{ericson2004real}. Although optimal solutions typically require complex algorithms, we find a simplified approach 
with linear running time to be good enough in our application: We perform the same separation operation as in {\em step2} but execute it twice: initially, we separate the polygons by moving them along the y-axis, and then repeat the process along the x-axis. We compare the two results and take the approach that results in the shortest total distance moved as the final choice. The parent node then adjusts to the average displacement of all child nodes. For instance, in Figure~\ref{fig:modifiedRT} {\em step3}, the two subtrees of node 1 are effectively separated by moving along the x-axis.

\begin{figure}
    \centering
    \includegraphics[width=1\linewidth]{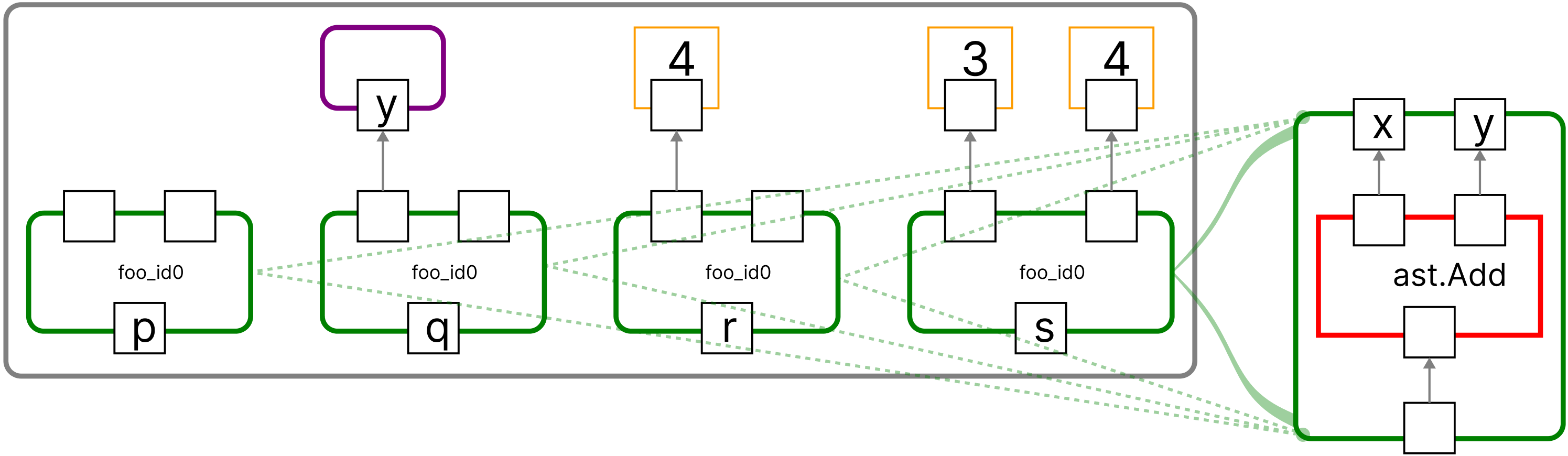}
    \captionsetup{belowskip=-10pt}
    \caption{Duplicate structures within one group are drawn with edges from the collapsed copies to the expanded form. In this application, we choose lightweight dashed edges to depict this relationship. Users of our layout can choose to suppress these edges and/or the expansion of duplicate siblings as layout parameters..}
    \label{fig:noDuplicate}
\end{figure}

We provide two methods for drawing duplicate members of the same group. They can be individually expanded or shown as a single subgroup with additional edges.
Figure~\ref{fig:noDuplicate} illustrates the latter case where all duplicate groups are connected to the expanded group with dashed lines, eliminating the need for drawing each expansion as required in the focus+context layout. This reduction of drawn elements is another benefit of the overview+detail approach. 

\section{Case Studies}
\label{sec:case}

We discuss the use of our layout for debugging and understanding compound networks generated during automated extraction of computations from source code. These case studies arose from deployed use of our layout by the author who was not involved in the design of the layout or the implementation but was involved in providing feedback towards a visualization that used the layout as its main view as well as in generating the datasets we used. We provide background for the use and visualization and then discuss how it was used in practice.

\begin{figure}
    \centering
    \includegraphics[width=1\linewidth]{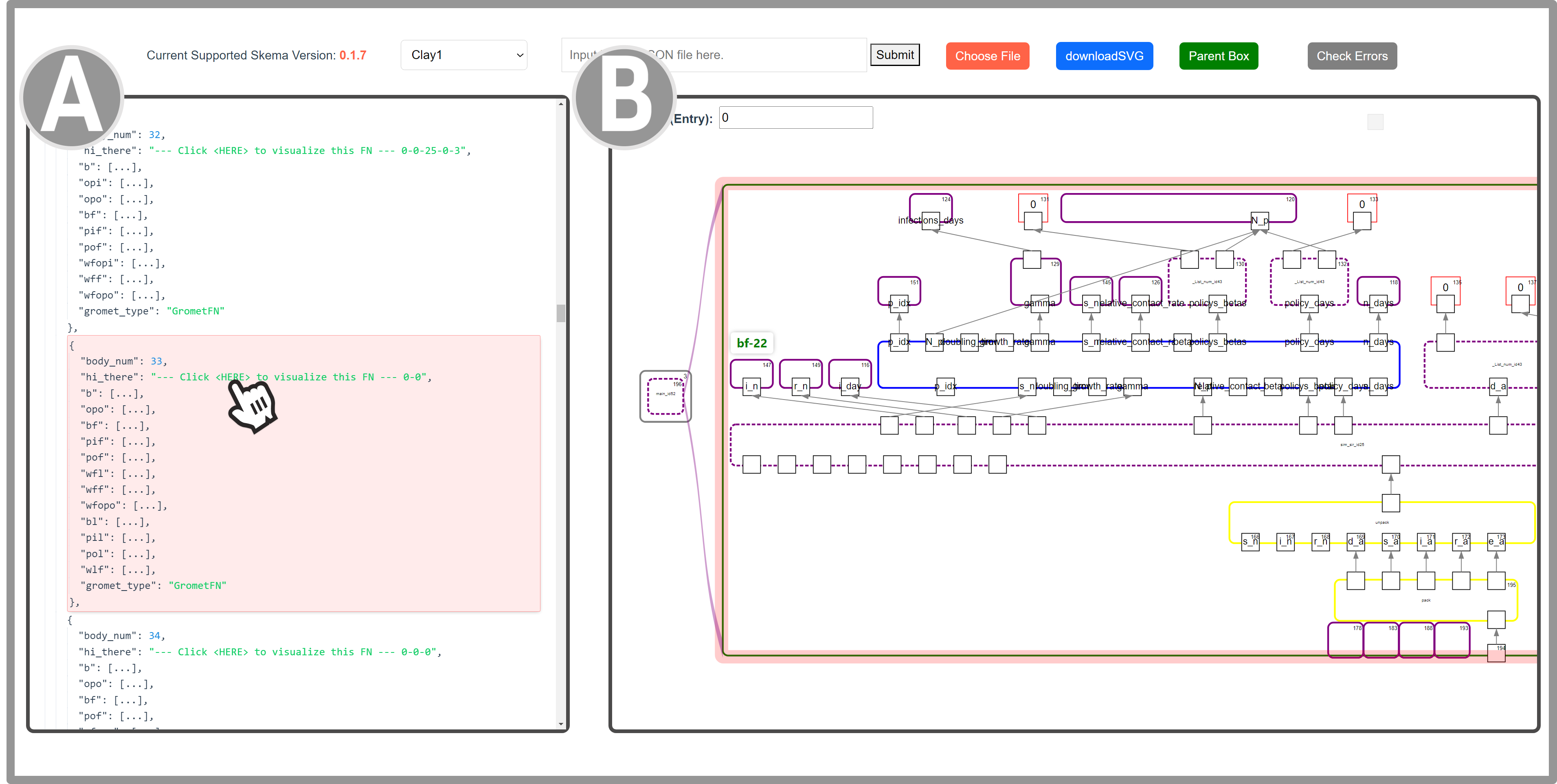}
    \caption{The interface used in the case study has two linked views. View A shows the raw JSON representing the Function Network. View B is the main view of our compound networks.}
    \label{fig:moviz}
\end{figure}

\subsection{Layout Application Background}
\label{sec:casebackground}

Our layout was used in the context of project involving automated extraction of computational models from source code and other documents. These models can then be analyzed either computationally or manually, combined with other models, and transpiled into new, more performant code. The models are stored in a custom abstract format, known as a Function Network~\cite{pyarelal2020automatesautomatedmodelassembly}, that can be cast as a compound network.

The author who performed these case studies is part of the extraction team. They were interested in the visualization to help them understand the Function Networks to help them create and debug steps further in the pipeline. As both the extraction and subsequent steps were in development, their use also led to identification of bugs in the extraction code.

The Function Networks encapsulate concepts like expressions, functions, loops, and modules as a generic `box' that can contain other boxes. Each box can have multiple input and output ports which represent values and can have names corresponding to variables extracted from the source. Links between ports indicate movement of values. In Function Networks, arrows point to a value's source rather than its destination. This abstraction is used by other analyses in the pipeline.

To match the illustrations in the project's documentation, we use categorical color to differentiate concepts like expressions and functions. We paired our layout with an auxiliary view showing the raw JSON representing the Function Network (\autoref{fig:moviz}). These  linked views enable seeking a particular box and expand the visualization such that the internals of that box are visible. 

\begin{figure}
    \centering
    \includegraphics[width=1\linewidth]{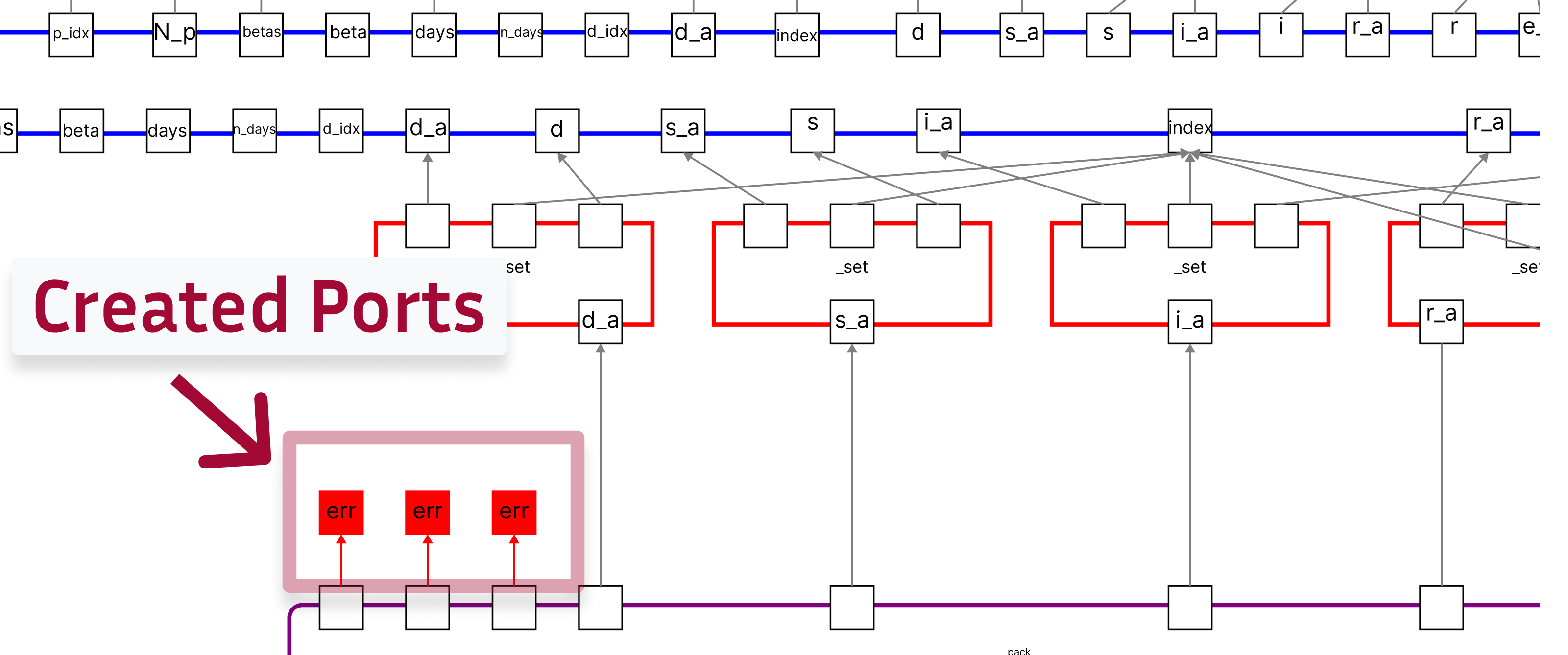}
    \caption{Undefined ports in the Function Network often signify bugs. We add them if referenced in red to make them more salient.}
    \label{fig:redports}
\end{figure}

A common bug in the Function Network extraction is referencing an undefined port. In this case, we created the port in our layout but drew it in solid red to indicate it was our creation (\autoref{fig:redports}).

\subsection{Use of Our Layout}
\label{sec:caseuses}

These case studies were extracted from brief notes the performing author wrote in a single month's summary report from the entire project. The first author followed up with the performing author for more detail. Both the report and follow up occurred during design of the multi-view visualizaiton, before publication was discussed.

\vspace{1ex}
\noindent\textbf{Using Our Layout for Debugging.} The author reported two instances in which our layout helped with debugging. In the first, they were able to determine a downstream error in running their code was caused by an upstream error in defining the Function Network. They noticed the solid red ports indicating missing definitions in the Function Network. Due to the placement in the layout, they were able to narrow down what part of the extraction process it was associated with. They said, ``With the way the ports are laid out, it’s easy to tell what kind of wires these are,'' referring to the implicit purpose of the links.

In the second instance, the author debugged their own code which makes inferecnes about the computation performed by the Function Network. They had made an assumption about the way the Function Network represented vectors. Once they checked the visualization, they understood their error, as well as the correct representation, and fixed their code. They particularly noted the ``cascading'' style of the layout helped them: ``If you see it, you’ll see a lot of those set operations going on. It’s kind of that cascading, from the [internal name] visualization.''

\vspace{1ex}
\noindent\textbf{Using Our Layout for Understanding.} The author also used our layout for understanding how Function Networks represent functions as parameters to other functions. The author was aware that other team members had added new specifications regarding function parameters. They used the layout with several test cases to understand the new specifications, often tweaking the code and examining the changes. Prior to the introduction of the layout, they would rely on documentation alone, which had a few hand-drawn examples of small, isolated uses. They said, ``It allowed me again, to get a better understanding of the structure and pretty quickly be able to update the code on my side.''

\section{Conclusion and Future Work}
We have presented an overview+detail layout for compound graphs and demonstrated its utility through case studies in debugging compound graphs representing computation. The layout specifically prioritizes displaying both the structure between higher level groups as well as the detail within groups, placing the two depictions, expanded and collapsed, near each other in the overall layout.

The impetus of our layout design was aiding in tracing through compound graphs representing processes that can include nested sub-processes, such as those found in biological processes, chemical equations, and computational data flow. Thus, it is designed primarily for directed networks. Specifically, we apply a convention of placing inputs on one side and outputs on the other through our layout's explicit ports.

We recommend this layout be used when the compound structure has several levels of nesting. In shallow networks or applications where higher levels are not as important, the limitations of focus+context and/or semantic zooming approaches in showing the higher-level behavior would have little effect on the analysis. 
We do not claim that our method is a superior substitute for previous methods, but rather an alternative choice in scenarios where understanding multiple levels simultaneously is crucial. Ultimately, empirical studies would need to be performed to more definitively and precisely understand the suitability of these methods to various compound graph tasks.

We note our tree layout method restricts the layout direction to right and down, which may still result in some long edges in some cases when the parent group is large and many subgroups are expanded. We made this compromise in our layout design for better computational efficiency and a clearer overall direction (from top-left to bottom-right). Alternatives placement algorithms, including ones that conisder all directions, may further improve the overview+detail layout approach.
\chang{Another limitation of our approach, similar to the traditional RT algorithm, is that it may produce elongated layouts. To address this, we allow users to pan and zoom within our application, as shown in~\autoref{fig:moviz}. Future work could explore methods to partially collapse elongated nodes or investigate creating area-adaptive layouts similar to Misue's approach~\cite{Misue2024}, but with plural layout directions and varied node sizes.}

\acknowledgments{The work reported here was supported by the Defense Advanced Research Projects Agency (DARPA), under agreement HR00112290092, and the National Science Foundation under award IIS-2324465.
}

\bibliographystyle{abbrv-doi-hyperref}

\bibliography{template}
\end{document}